\documentclass[aps,pra,floatfix,twocolumn,showpacs,preprintnumbers,amsmath,amssymb,superscriptaddress]{revtex4-2}
\usepackage{graphicx}
\usepackage{bm}
\usepackage{color} 
\usepackage{braket}
\usepackage{epstopdf}

\begin{document}
\title{Designing Quantum States with Tunable Magnetic Textures}
\author{Siphiwo R. Dlamini}
\author{Alex Matos-Abiague}
\affiliation{Department of Physics and Astronomy, Wayne State University, Detroit, MI 48201, USA}

\date{\today}
 
\begin{abstract}
We theoretically investigate the effects of tunable magnetic fringe fields generated by arrays of switchable magnetic junctions (MJs) on the quantum states of an underlying two-dimensional (2D) system formed in a semiconductor quantum well. The magnetic landscape generated by the MJ-array can be reconfigured on the nanometer scale by electrically switching the magnetic state of individual MJs. The interaction of carriers’ spin with the fringe fields generates effective spin-dependent potentials acting like quantum dots (barriers) for carriers with spin parallel (antiparallel) to the magnetic texture. The position and depth (height) of the magnetically generated quantum dots (barriers), as well as the coupling between them, can be tuned by modulating the magnetic texture through switchings of individual MJs. This enables the magnetic control, manipulation, and design of quantum states, their spin, and transport properties.
\end{abstract}

\maketitle

\section{Introduction}

Electrical gating is a well established method for controlling electronic states. Typical examples are the modulation of potential barriers and the creation of quantum dots (QDs) by using gate voltages. Alternatively, the use of tunable magnetic fields for controlling and designing quantum states and their charge and spin properties may provide new functionalities for developing new electronic/spintronic devices. The magnetic field couples to the electron motion through the kinetic momentum, and to the electron spin through the Zeeman interaction, which can be sizable for materials with a large effective $g$-factor, even at relatively small magnetic fields. However, the applied magnetic fields are usually generated by charge currents or by external magnets with no possibility of modulating the field at the nanoscale, and limited use in quantum devices. A possible way to overcome this limitation is to use the fringe fields generated by switchable nanomagnets.

The fringe fields generated by tiny magnets ranging from few microns down to the nanometer scale have been experimentally proven to produce sizable effects on the transport properties of nanosystems \cite{Betthausen2012:S,Belkin2006:APL}.  This has prompted various experimental efforts to use fringe fields for designing micro and nanodevices whose functionalities vary from magnetometry \cite{Kubrak1999:APL}, magnetic sensing \cite{Heremans1990:APL,Kanda2014:APL}, rectification \cite{Szelong2019:JAP}, quantum Hall edge states sensing \cite{Karmakar2011:PRL}, among others. The feasibility of using the fringe fields of magnetic junctions (MJs) to locally modulate a magnetic texture and change the transport properties of a (Cd,Mn)Te quantum well has been demonstrated in the realization of a spin-transistor \cite{Betthausen2012:S}.

\begin{figure}[t]
	\centering
	\includegraphics[width=1\linewidth]{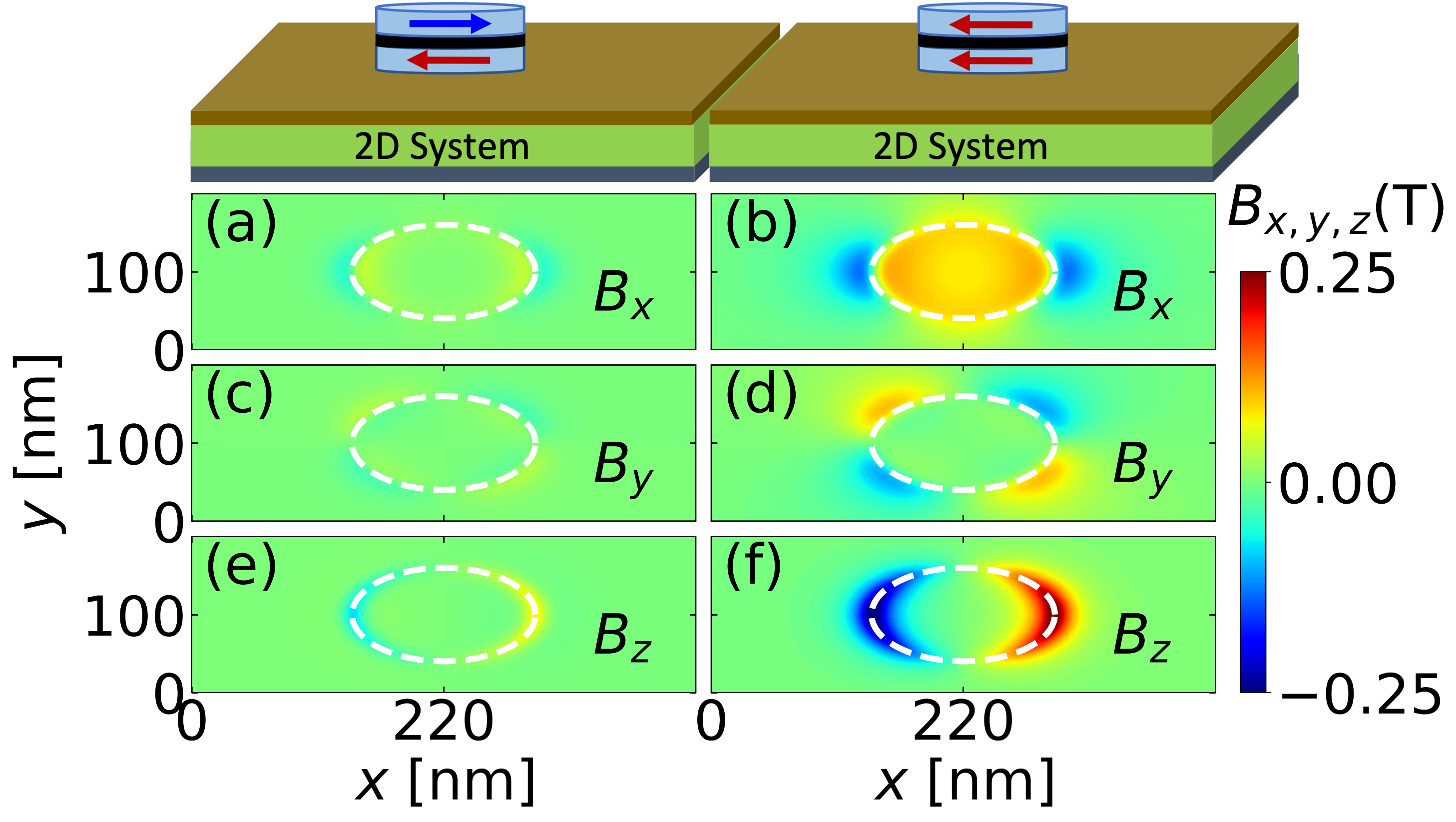}
	\caption{Fringe fields generated by a single MJ on a 2D system underneath. Left column: (a), (c), and (e) show the position dependence of the magnetic field components on the plane of the 2D system when the MJ is in the antiparallel (OFF) configuration. Right column: same as left column but for an MJ in the parallel (ON) configuration. Note that the fields in the left column (MJ is OFF) are much smaller than those in the right column (MJ is ON). Dashed-line contours represent the projection of the MJ elliptical shape on the 2D system.}
	\label{fig:fields}
\end{figure}

Here we provide a theoretical investigation of the effects of tunable magnetic textures generated by arrays of switchable, nanometric magnetic junctions (MJs) on the quantum properties of a two-dimensional (2D) system formed in a semiconductor quantum well underneath. As illustrated in Fig.~\ref{fig:fields}, the strength of the fringe field generated by an MJ strongly depends on whether the junction is in the parallel (ON state) or antiparallel (OFF state) configuration. Hence, the magnetic texture generated by an MJ-array can be locally modulated by switching the magnetic state of individual MJs. This allows for creating non-volatile magnetic textures that can be reconfigured on the nanometric scale. 

The interaction of the carriers with the magnetic texture leads to changes in the quantum properties of the underlying 2D system. The bottom layers of the MJs are assumed to be fixed, while the top layer can be switched by passing line currents and using the spin-orbit torque (SOT) generated by the spin Hall and/or Rashba effects \cite{Liu2012:S,Fukami2016:NN,Manchon2009:PRB,Matos-Abiague2009:PRB,Manchon2015:NM}. SOT-induced switching can be performed within a time scale in the nanoseconds down to the sub-nanosecond regime \cite{Jhuria2020:NE,Garello2014:APL,Cubukcu2018:IEEETM,Krizakova2020:APL} with low power consumption \cite{Jhuria2020:NE,Garello2014:APL,Cubukcu2018:IEEETM,Krizakova2020:APL,Kent2015:NN,Endoh2018:JLPEA}. Optical switching of MJs using short laser pulses has also been demonstrated \cite{Chen2017:PRApp,Yang2017:SA}. Reconfigurable magnetic textures generated by MJ-arrays have previously been proposed as possible control knobs for the creation and manipulation of Majorana bound states in topological superconductors \cite{Matos-Abiague2017:SSC,Fatin2016:PRL,Zhou2019:PRB}.

\section{Theoretical Model}

The Hamiltonian describing the states of the carriers in the 2D system under the magnetic fringe field $\mathbf{B}(\mathbf{r})$ is given by
\begin{equation}\label{Hdef}
H=\frac{\boldsymbol{\pi}^2}{2m^\ast}-\frac{g^\ast\mu_B}{2}\;\mathbf{B}(\mathbf{r})\cdot\boldsymbol{\sigma},
\end{equation}
where $\boldsymbol{\pi}=\mathbf{p}-e\mathbf{A(\mathbf{r})}$ is the kinetic momentum and $\mathbf{A}(\mathbf{r})$ is the vector potential generating the magnetic texture $\mathbf{B}(\mathbf{r})=\boldsymbol{\nabla}\times\mathbf{A}(\mathbf{r})$. The last term is the Zeeman interaction with $g^\ast$ denoting the effective $g$-factor, and $\mu_B$ and $\boldsymbol{\sigma}=(\sigma_x, \sigma_y, \sigma_z)$ representing the Bohr magneton and the vector of Pauli matrices, respectively.

Using spherical coordinates, a general magnetic texture can be written as,
\begin{equation}
	\mathbf{B}(\mathbf{r}) = |\mathbf{B}(\mathbf{r})| \begin{pmatrix}
		\sin \theta(\mathbf{r}) \cos \phi(\mathbf{r})\\
		\sin\theta(\mathbf{r}) \sin \phi(\mathbf{r})\\
		\cos\theta(\mathbf{r})
	\end{pmatrix}.
\end{equation}

\subsection{Bound States and Synthetic SOC}

To understand the origin of the bound states produced by the magnetic texture,  we first diagonalize the Zeeman interaction by performing local spin rotations aligning the spin quantization axis to the local magnetic field direction. Such a rotation is given by the unitary transformation,
\begin{equation}
U(\mathbf{r}) = \begin{pmatrix}
e^{-i\phi(\mathbf{r})/2} \cos\frac{\theta(\mathbf{r})}{2} & - e^{-i\phi(\mathbf{r})/2}\sin\frac{\theta(\mathbf{r})}{2} \\
e^{i\phi(\mathbf{r})/2} \sin\frac{\theta(\mathbf{r})}{2} & e^{i\phi(\mathbf{r})/2}\cos\frac{\theta(\mathbf{r})}{2}
\end{pmatrix}\;.
\end{equation}
For simplicity of the notation, the position dependence of the spherical angles will be omitted.

In the rotated frame, the Hamiltonian reads as
\begin{equation}\label{h-rotated}
H'=U^\dagger H U
= \frac{\left[\boldsymbol{\pi}-e\boldsymbol{\mathbb{A}}(\mathbf{r})\right]^2}{2m^*}- \frac{g^\ast\mu_B}{2}|\mathbf{B}(\mathbf{r})| \sigma_z\;,
\end{equation}
where,
\begin{equation}\label{non-abelian-field}
\boldsymbol{\mathbb{A}}(\mathbf{r}) = \frac{\phi_0}{2\pi} \left(-\sigma_x \boldsymbol{\nabla}\phi \sin \theta + \sigma_y \boldsymbol{\nabla} \theta + \sigma_z \boldsymbol{\nabla}\phi\cos\theta\right),
\end{equation}
and $\phi_0 = h/(2e)$ is the magnetic flux quantum.

Expanding the kinetic term yields,
\begin{equation}\label{h-rotated-expanded}
H' = \frac{\boldsymbol{\pi}^2}{2m^*}+V_B(\mathbf{r})+H_{\rm sso}-  \frac{g^\ast\mu_B}{2}|\mathbf{B}(\mathbf{r})| \sigma_z\;,
\end{equation}
where the synthetic SOC induced by the non-Abelian field $\boldsymbol{\mathbb{A}}(\mathbf{r})$ reads
\begin{equation}\label{hsso}
H_{\rm sso}= -\frac{e}{2 m^\ast}\left[\boldsymbol{\mathbb{A}}(\mathbf{r})\cdot\boldsymbol{\pi} + {\rm h.c.}\right]\;,
\end{equation}
and
\begin{equation}
    V_B(\mathbf{r})=e^2\frac{\boldsymbol{ \mathbb{A}}(\mathbf{r})\cdot \boldsymbol{ \mathbb{A}}(\mathbf{r})}{2m^{\ast}} =\hbar^2\frac{\left(|\boldsymbol{\nabla}\phi|^2+|\boldsymbol{\nabla}\theta|^2\right)}{8m^{\ast}}.
\end{equation}

Unlike the real, electric-field induced SOC (e.g., atomic SOC, Rashba SOC, etc), the effective synthetic SOC in Eq.~(\ref{hsso}) originates from the noncollinearity of the magnetic texture. The strength, form, symmetry, and spatial range of the synthetic SOC is not determined by the atomic structure of the system but by the pace of the fringe field direction changes. Therefore, the properties and symmetry of the synthetic SOC can be designed and controlled by properly tuning the magnetic texture generated by the array of MJs.

In general, due to the complexity of the fringe fields generated by MJ arrays, the synthetic SOC must be computed numerically. However, one can obtain analytical expressions for the case of simpler magnetic textures. For example, in the case of a helical texture with wavelength $\lambda$ and
\begin{equation}\label{helical-j}
\mathbf{B}(\mathbf{r}) = B\begin{pmatrix}
\sin (2\pi x/\lambda)\\
0 \\
\cos(2\pi x/\lambda)
\end{pmatrix},
\end{equation}
the synthetic SOC induced by the magnetic helix acquires the form
\begin{equation}
H_{\rm sso}=-\frac{2\pi\hbar}{m^\ast \lambda}\pi_x\sigma_y\;,
\end{equation}
which resembles the Rashba SOC in a quantum wire along the $\hat{\mathbf{x}}$ direction \cite{Zutic2004:RMP,Winkler2003,Fabian2007:APS}. Note that the strength of the synthetic SOC is inversely proportional to the wavelength of the texture. Therefore, for the synthetic SOC to be sizable, one typically needs textures with characteristic lengths of variation on the order of few hundred nanometers or shorter.

When the fringe field is large enough, the spin-flip transitions are energetically unfavorable, and the spin of the carrier tends to follow the direction of the texture. This adiabatic regime occurs when the rate at which the local spin quantization axis varies is much slower than the spin precession frequency \cite{Bruno2004:PRL}. The rate of magnetic field variation experienced by the carriers at the Fermi level is determined by $v_F/\lambda$, where $v_F$ and $\lambda$ are the Fermi velocity and the characteristic length of variation of the spin quantization axis, respectively. The spin precession frequency can be approximated as $\omega_s\approx g^\ast \mu_B \bar{B}/\hbar$, where $\bar{B}$ represents the field magnitude average over the characteristic variation length. The adiabaticity condition can then be written as \cite{Bruno2004:PRL},
\begin{equation}
    \frac{\hbar v_F}{\lambda}\ll g^\ast \mu_B \bar{B}
\end{equation}

In the adiabatic regime, the rotated Hamiltonian, $H'$, becomes diagonal in the spin degree of freedom, and reduces to,
\begin{equation}
    H'=\frac{\left[\mathbf{p}-e\mathbf{A}_{\sigma}(\mathbf{r})\right]^2}{2m^*}+V_{\sigma},
\end{equation}
where,
\begin{equation}
    \mathbf{A}_{\sigma}(\mathbf{r})=\mathbf{A}(\mathbf{r})+\sigma\frac{\phi_0}{2\pi}\boldsymbol{\nabla}\phi \cos\theta,
\end{equation}
and
\begin{equation}\label{v-sigma-1}
    V_{\sigma}(\mathbf{r})=\hbar^2\frac{\left(|\sin\phi\boldsymbol{\nabla}\phi|^2+|\boldsymbol{\nabla}\theta|^2\right)}{8m^{\ast}}-\sigma \frac{g^\ast\mu_B}{2}|\mathbf{B}(\mathbf{r})|,
\end{equation}
are effective spin-dependent potentials. The spin quantum number, $\sigma=1$ ($\sigma=-1$) indicates that at any position, the spin is parallel (antiparallel) to the magnetic field (assuming $g^\ast>0$).

The spin-dependent part of the vector potential $\mathbf{A}_\sigma$ generates the effective Lorentz force responsible for the topological Hall effect \cite{Bruno2004:PRL}, while the spin-dependent potential $V_\sigma$ may lead to the formation of bound states in systems with large $g$-factor, where the second contribution in Eq.~(\ref{v-sigma-1}) dominates \cite{Note:1}. In such a case, one can approximate the spin-dependent potential as,
\begin{equation}\label{v-sigma-2}
    V_{\sigma}(\mathbf{r})\approx-\sigma \frac{g^\ast\mu_B}{2}|\mathbf{B}(\mathbf{r})|.
\end{equation}
For $\sigma =1$ the spin-dependent potential, $V_{\sigma=1}(\mathbf{r})<0$, becomes attractive over some finite region(s) of the 2D system and bound states may emerge \cite{Note:2}. This open the possibility for using the textures generated by MJ-arrays as reconfigurable magnetic gates enabling the design and manipulation of quantum states without involving electrostatic potentials.

\section{Numerical Approach}

\subsection{Fringe Fields}

For the numerical simulations we considered MJs composed of a 2~nm thin Si layer sandwiched between two ferromagnetic, CoFe layers of thickness 7~nm each. The magnetization saturation of CoFe was set to $M_s = 1.7\times 10^6$~A/m. The MJs have elliptical cross-sections with symmetry axes of lengths 160~nm and 120~nm in the $x$ and $y$ directions respectively and have a 40~nm separation from each other. The fringe fields are calculated on the plane of the 2D system (located 10~nm below the MJs) by using JOOMMF \cite{Beg2017:AIPA,joommf}, an interface integrating the micromagnetic simulation package OOMMF \cite{oommf} with Python and Jupyter Notebook. A discretized grid with lattice spacing of 1~nm was used for the fringe field calculations.

\subsection{Quantum States of 2D System}

\subsubsection{Spectral Properties}

In the case of a two-dimensional electron gas with confinement in the $z$ direction and in small magnetic fields, the orbital motion of the carriers is only affected by the $z$-component of the magnetic field. Within the tight-binding approximation, we discretize the system Hamiltonian on a square lattice with lattice parameter $a$. The position of the lattice points, $(x, y)=\mathbf{n} a$, are characterized by the vector $\mathbf{n}=(n_x,n_y)$, where $n_x$ and $n_y$ are integers. The tight-binding Hamiltonian can then be written as, 

\begin{eqnarray}\label{h-tb}
H_{\rm TB} &=& \sum_{\textbf{n},\sigma}(4t-\mu)|\textbf{n},\sigma\rangle\langle \textbf{n},\sigma|-\left[\sum_{\langle \textbf{n},\textbf{n}^\prime \rangle,\sigma} t_{\mathbf{n' n}}|\textbf{n},\sigma\rangle\langle \textbf{n}^\prime,\sigma|\right.\nonumber\\
&-& \left.\frac{g^{\ast}\mu_B}{\hbar}\sum_{\textbf{n},\sigma,\sigma^\prime} (\textbf{B}_\textbf{n} \cdot \textbf{S}_{\sigma\sigma^\prime}) |\textbf{n},\sigma\rangle \langle\textbf{n},\sigma^\prime|+{\rm H.c.}\right],
\end{eqnarray}
where $\langle \mathbf{r}|\mathbf{n},\sigma\rangle=\psi_\sigma(\mathbf{n}a)$ is the wave function, $t=\hbar^2/(2m^\ast a^2)$ is the hopping parameter, $\mathbf{B}_\textbf{n}$ is the magnetic field at site $\textbf{n}$, $\textbf{S}_{\sigma\sigma^\prime}$ are the spin matrix elements, with components $S^{x,y,z}_{\sigma,\sigma^\prime}=\langle \sigma|\sigma_{x,y,z}|\sigma^\prime \rangle \hbar/2$, and $\langle \textbf{n},\textbf{n}^\prime \rangle$ indicates the sum is over nearest neighbors. The magnetic orbital effects are taken into account by performing the Peierls substitution,
\begin{equation}\label{t-peierls}
   t_{\mathbf{n' n}}=t\; e^{-i\frac{e}{\hbar}\int_{\mathbf{n}}^{\mathbf{n}^\prime}\mathbf{A}\cdot d\mathbf{l}},
\end{equation}
where $\int_{\mathbf{n}}^{\mathbf{n}^\prime}\mathbf{A}\cdot d\mathbf{l}$ is the integral of the vector potential along the hopping path, from site $\mathbf{n}$ to $\mathbf{n'}$.

The energy spectrum, eigenspinors, and local density of states (LDOS) are computed by using the Kwant package \cite{Groth2014:NJP}.

\subsubsection{Transport Properties}

For the quantum transport properties, semi-infinite metallic leads are attached on either sides of the scattering region along the $x$ direction. The zero-temperature differential conductance, $G$, is computed by using Landauer's formula,
\begin{equation}
    G=\frac{e^2}{h}\sum_{k\sigma\in R,l\sigma'\in L}|\mathcal{S}_{k\sigma,l\sigma'}|^2,
\end{equation}
as implemented in the Kwant package \cite{Groth2014:NJP}. Here $\mathcal{S}_{k\sigma,l\sigma'}$ denotes the scattering matrix element corresponding to the conducting channels $k\sigma$ and $l\sigma'$ in the right ($R$) and left (L) leads.

\section{Results}

Numerical calculations were performed for a (Cd,Mn)Te quantum well with effective $g$-factor, $g^\ast = 300$, and effective mass $m^\ast=0.1~m_0$ (with $m_0$ as the bare electron mass). The lattice constant was set to $4$~nm, which corresponds to a hopping energy, $t=23.8$~meV.

\subsection{Single MJ}

\begin{figure}[t]
\centering
\includegraphics[width=1\linewidth]{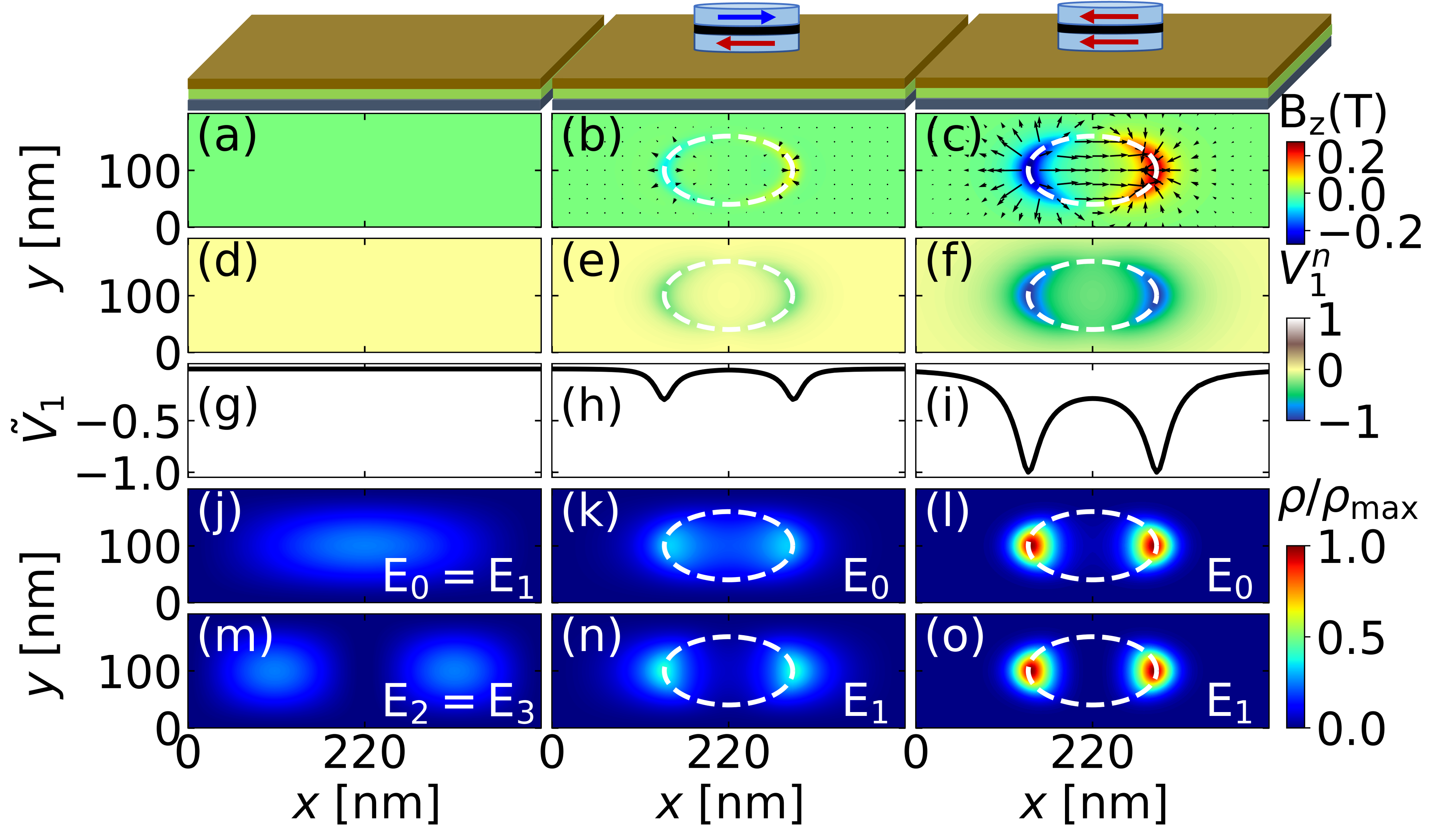}
\caption{Magnetic textures generated by a single MJ and their effects on the quantum states of the 2D system underneath. Each column shows results for the corresponding system configurations sketched at the top of the figure. (a)-(c) Position dependence of the magnetic fringe fields. Colors indicate the strength of the field component along the $z$ direction and the arrows correspond to the in-plane components. (d)-(f) Position dependence of the normalized spin-dependent potential, $V_{1}^n=V_{1}/(\max|V_{1}|)$, where $\max|V_{1}|=2.39$~meV is taken over all positions and configurations. (g)-(i) Variation of the normalized spin-dependent potential along the $x$ direction for $y=0$ [$\tilde{V}_1(x)=V_1^n(x,0)$]. (j)-(l) and (m)-(n) Probability densities (normalized to their maximum value) for the ground and first excited states, respectively.}
\label{fig:rho-1}
\end{figure}

We considered a 2D system with size of 440~nm and 200~nm along the $x$ and $y$ directions, respectively. The results of numerical simulations illustrating the emergence of bound states is shown in Fig.~\ref{fig:rho-1} for the cases of none and a single MJ acting on the underlying 2D system. The left column corresponds to the system in the absence of fringe fields, while the middle and right columns correspond to the cases of a single MJ in the OFF (antiparallel) and ON (parallel) configurations, respectively. The fringe fields are shown Figs.~\ref{fig:rho-1}(a), (b), and (c), where colors represent the field strengths in the $z$ direction (perpendicular to the plane) and the arrows correspond to the in-plane field components. The position dependence of the spin-dependent potential generated by the fringe fields [see Eq.~(\ref{v-sigma-2})], $V_{\sigma =1}$ is displayed in Figs.~\ref{fig:rho-1} (d)-(f) for $\sigma =1$. To facilitate the comparison between the different configurations, the spin-dependent potential has been normalized to the maximum of its absolute value ($|V_{\sigma = 1}|_{\rm max}=2.39$~meV). In regions where the fringe fields are strong, the spin-dependent potential becomes attractive for carriers with $\sigma =1$, forming effective quantum dots (QDs). The traces of the normalized spin-dependent potential along the $x$ direction and $y=0$ [see Figs.~\ref{fig:rho-1}(g-(i)] illustrate the confining potential giving rise to the formation of fringe-field generated effective double QDs. As discussed below, the effective QDs can lead to the emergence of bound states.

The probability density of the ground and first excited states of the 2D system in the absence of fringe fields is shown in Figs.~\ref{fig:rho-1}(j) and (m). In this case the states are spin degenerate, with energies $E_0=E_1=0.11$~mev and $E_2=E_3=0.16$~meV. In the absence of fringe fields, the finite-size 2D system behaves as a rectangular quantum box with a spin-degenerate ground state with a probability density spread over the central region of the box and the first-excite state probability density exhibiting two maxima. The fringe field generated by an MJ in the OFF configuration is relatively small [see Fig.~\ref{fig:rho-1}(b)] but enough to break the spin degeneracy. However, the effective QDs generated by the spin-dependent confining potential are too shallow [see Fig.~\ref{fig:rho-1}(e) and (h)] to support bound states, and the probability density of the lowest-energy states (with $E_0=0.01$~meV and $E_1=0.04$~meV) is mainly determined by the confining potential of the rectangular quantum box defining the 2D system, as shown in Figs.~\ref{fig:rho-1}(k) and (n). When the MJ is switched to the ON configuration, the spin-dependent potential becomes strong enough and quasi-degenerate bound states (with $E_0=-1.06$~mev and $E_1=-1.04$~meV) localized in the effective coupled QDs [see Figs.~\ref{fig:rho-1}(f) and (i)] emerge, as illustrated in Figs.~\ref{fig:rho-1}(l) and (o).

\begin{figure}[t]
\centering
\includegraphics[width=1\linewidth]{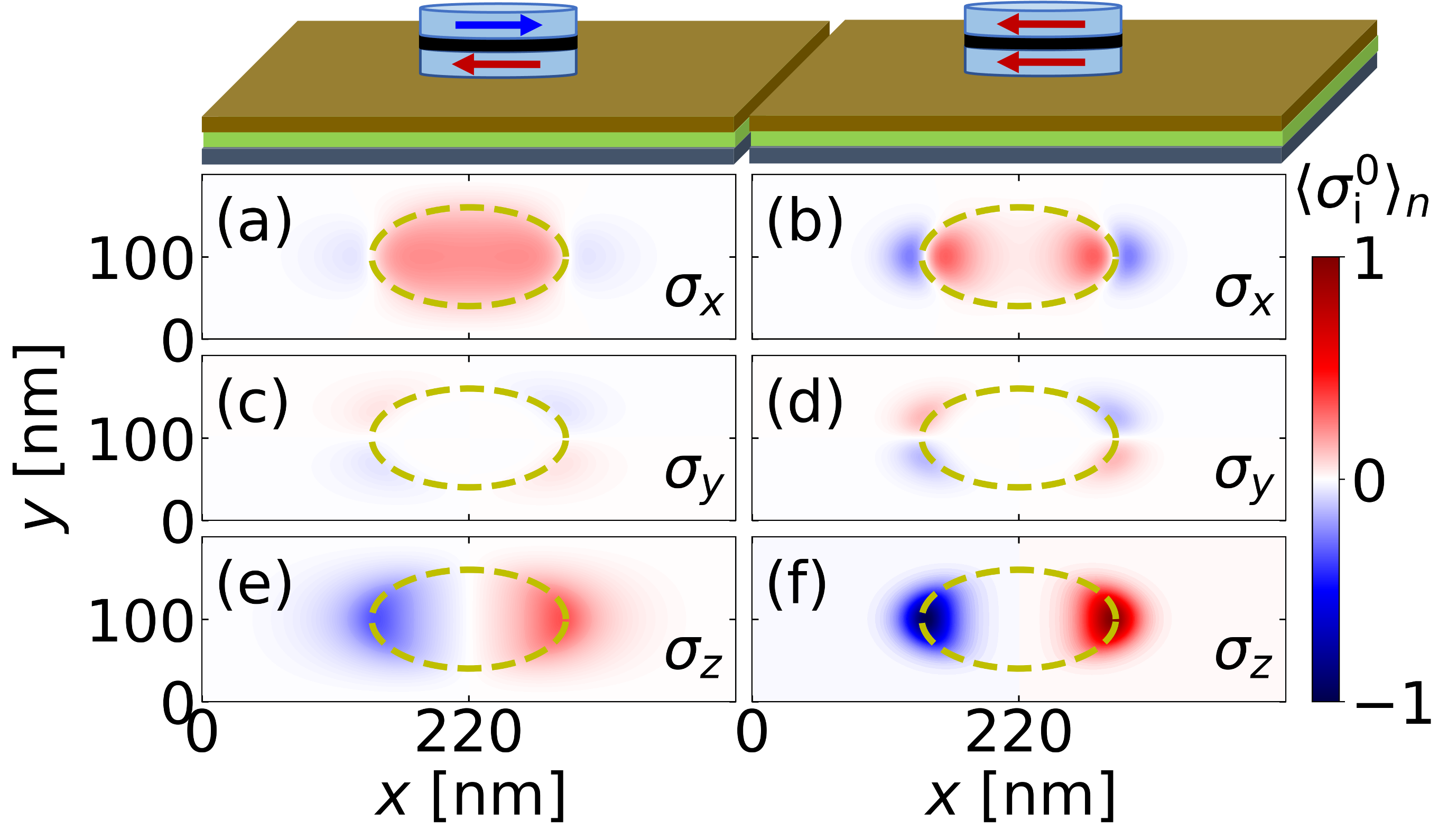}
\caption{Position dependence of the ground-state spin density (normalized to its maximum value) generated by the fringe fields of an MJ in the OFF (left column) and ON (right column) states.}
\label{fig:s-1-0}
\end{figure}

The ground-state spin density (normalized to its maximum value) generated by the fringe field of a single MJ is shown in Fig.~\ref{fig:s-1-0}. A clear correlation between the spin density behavior and the magnetic field generated by the MJ can be clearly seen by comparing Figs.~\ref{fig:fields} and \ref{fig:s-1-0}. In particular, the symmetry of the spin density components is directly related to the symmetry properties of the fringe fields generated by the MJ,
\begin{equation}\label{inv-x}
[\mathcal{R}_x,B_x]=0\;,\;\{\mathcal{R}_x,B_y\}=\{\mathcal{R}_x,B_z\}=0,
\end{equation}
and
\begin{equation}\label{inv-y}
[\mathcal{R}_y,B_x]=[\mathcal{R}_y,B_z]=0\;,\;\{\mathcal{R}_y,B_y\}=0, 
\end{equation}
where square and curly brackets denote commutators and anticommutators, respectively, and the operator $\mathcal{R}_x$ ($\mathcal{R}_y$) represents inversion in position space with respect to the $x$ ($y$) axis (assuming the origin of coordinates is set at the center of the 2D system). The relations in Eq.~(\ref{inv-x}) leads to the Hamiltonian symmetry,
\begin{equation}\label{h-sim-rx}
    [H,\mathcal{R}_x\mathcal{R}_{sx}]=0,
\end{equation}
Similarly, Eq.~(\ref{inv-y}) leads to,
\begin{equation}\label{h-sim-ry}
    \{H,\mathcal{R}_y\mathcal{R}_{sy}\}=0.
\end{equation}
In the equations above, $\mathcal{R}_{sx}$ ($\mathcal{R}_{sy}$) is the operator of spin reflection with respect to the plane containing the coordinate origin and perpendicular to the $x$ ($y$) axis. As a direct consequence of Eq.~(\ref{h-sim-rx}), the spin-density components must obey the symmetry relations,
\begin{eqnarray}\label{s-sim-x}
    \langle\sigma_x\rangle(x,y)&=&\langle\sigma_x\rangle(-x,y)\;,\nonumber\\
    \langle\sigma_y\rangle(x,y)&=&-\langle\sigma_y\rangle(-x,y)\;,\nonumber\\
    \langle\sigma_z\rangle(x,y)&=&-\langle\sigma_z\rangle(-x,y),
\end{eqnarray}
while Eq.~(\ref{h-sim-ry}) yields,
\begin{eqnarray}\label{s-sim-y}
    \langle\sigma_x\rangle(x,y)&=&\langle\sigma_x\rangle(x,-y)\;,\nonumber\\
    \langle\sigma_y\rangle(x,y)&=&-\langle\sigma_y\rangle(x,-y)\;,\nonumber\\
    \langle\sigma_z\rangle(x,y)&=&\langle\sigma_z\rangle(x,-y).
\end{eqnarray}
Figure~\ref{fig:s-1-0} confirms that the spin-density components indeed fulfill these symmetry relations.

Equations~(\ref{h-sim-rx}) and (\ref{s-sim-x}) remain valid when the effect of Rashba SOC is taken into account but no longer hold in the presence of Dresselhaus SOC. Conversely, Dresselhaus SOC preserves the symmetry relations in Eqs.~(\ref{h-sim-ry}) and (\ref{s-sim-y}) but Rashba SOC breaks them (see Appendix for more details).

\begin{figure}[t]
\centering
\includegraphics[width=1\linewidth]{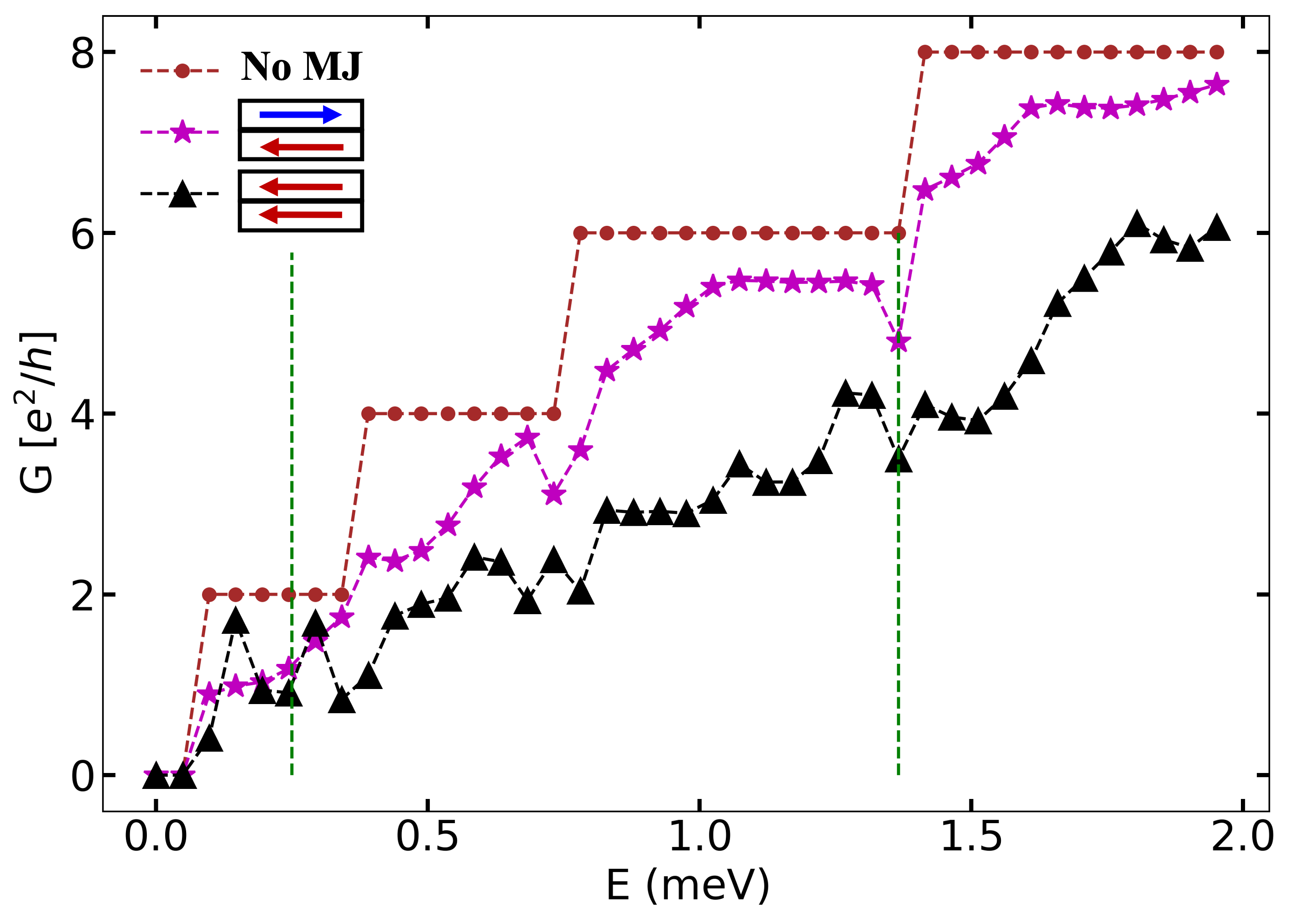}
\caption{Energy dependence of the differential conductance for different MJ configurations. Vertical dashed lines indicate two of the energy values at which the conductance has local minima when the MJ is ON. These energy values were used to compute the LDOS in Fig.~\ref{fig:ldos-1}.}
\label{fig:g-1}
\end{figure}

The emergence of bound states and the changes in the states localization induced by the magnetic textures leads to strong dependence of the charge transport on the magnetic configuration of the MJ. This results in sizable differences in the energy dependence of the differential conductance when the MJ is switched between the OFF and ON states, as shown in Fig.~\ref{fig:g-1}.

In general the spin-dependent potentials $\mathbf{A}_\sigma$ and $V_\sigma$ affect the transmission through the 2D system. In particular, $V_\sigma$ is attractive for states with $\sigma=1$, while carriers with $\sigma=-1$ perceive the spin-dependent potential $V_{-1}=-V_1$ as potential barriers. Therefore, the transmission probability is expected to strongly depend on the carrier energy and the strength, variation, and location of the attractive QDs and barriers. This results in sizable differences in the energy dependence of the differential conductance when the MJ is switched between the OFF and ON states, as shown in Fig.~\ref{fig:g-1}. In the absence of magnetic fringe fields the differential conductance exhibits quantized plateaus, as expected for ballistic transport, where each available channel at a given energy contributes with a conductance quantum $e^2/h$. Since at zero magnetic field the channels are spin degenerate, the conductance jumps between even numbers of $e^2/h$. In the presence of fringe fields the transmission becomes energy-dependent and the conductance is no longer quantized.

When the MJ is in the OFF state, the field-generated spin-dependent potential forms a relatively weak double potential barrier for carriers with $\sigma=-1$ [this follows from the relation $V_{-1}=-V_1$ and Figs.~\ref{fig:rho-1}(e) and (h)], resulting in an overall decrease in the transmission probability and conductance, compared to the case of zero field (see Fig.~\ref{fig:g-1}). A similar situation occurs when the MJ is in the ON state, but since the spin-dependent potential is stronger in this case, the overall decrease in the conductance becomes even larger (see Fig.~\ref{fig:g-1}). However, there are a couple of energy values at which the conductance for the MJ in the ON state approaches the zero-field conductance value of $2e^2/h$ (see the two conductance peaks in the energy interval between 0 and 0.5~meV). These conductance maxima originates from resonant tunneling through the effective double potential barrier generated by the fringe field. Conversely, the conductance decrease (see, for example, the local minima at energies $E=0.25$~mev and $E=1.37$~meV, indicated with vertical dashed lines in Fig.~\ref{fig:g-1}) can be correlated to the lack of connectivity in the local density of states (LDOS). 

\begin{figure}[t]
\centering
\includegraphics[width=1\linewidth]{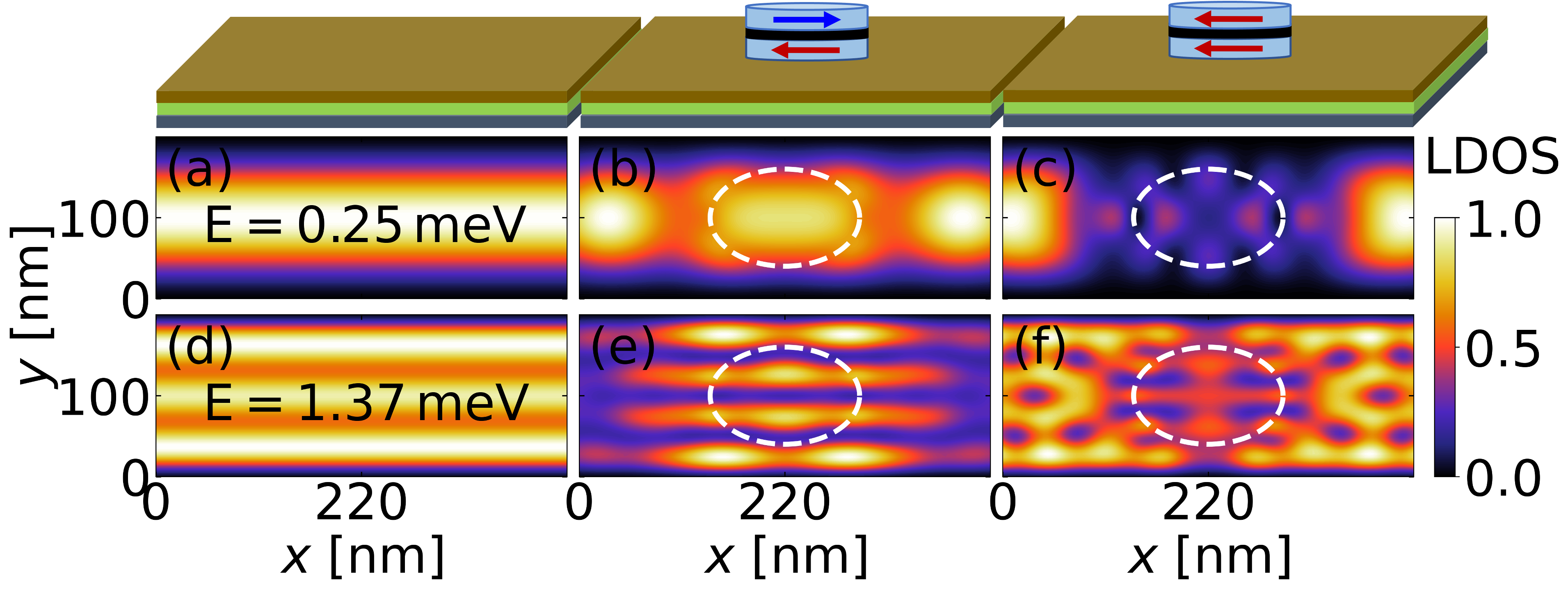}
\caption{Position dependence of the local density of states (normalized to its maximum value) at the energies indicated by dashed lines in Fig.~\ref{fig:g-1}, namely, $E=0.25$~meV [(a)-(c)] and $E=1.37$~meV [(d)-(f)].}
\label{fig:ldos-1}
\end{figure}

The LDOS normalized to its maximum value is shown in Figs.~\ref{fig:ldos-1}(a)-(c) and (d)-(f) for $E=0.25$~mev and $E=1.37$~meV, respectively. In the absence of fringe fields the spin-degenerate LDOS is constant along the transport direction ($x$ axis), providing high transmission channels connected all the way across the scattering region [see Figs.~\ref{fig:ldos-1}(a) and (d)]. However, in the presence of the magnetic textures, the LDOS is no longer homogeneous along the transport direction and the emergence of regions with low LDOS lowers the connectivity between high LDOS regions [see Figs.~\ref{fig:ldos-1}(c)-(d) and (e)-(f)], producing a decrease in the conductance.

\subsection{3-MJ Array}

As discussed above, the properties of the bound states formed in the 2D system depend on the specific form of the magnetic texture generated by the MJ fringe field, which can be tuned by changing the state of the MJ. Therefore, the use of arrays containing multiple MJs provides a way of designing and tuning a large number of different sets of quantum states. For example, an array containing $n$ MJs (each with two distinctive states, ON and OFF) can generate $2^n$ different magnetic textures and, therefore, $2^n$ sets of quantum states (each set composed of ground and corresponding excited states) with different energy spectra and localization properties. In dependence on the MJ switching time, one can create a coherence superposition of different quantum states or transit adiabatically from a quantum state to another.

\begin{figure}[t]
\centering
\includegraphics[width=1\linewidth]{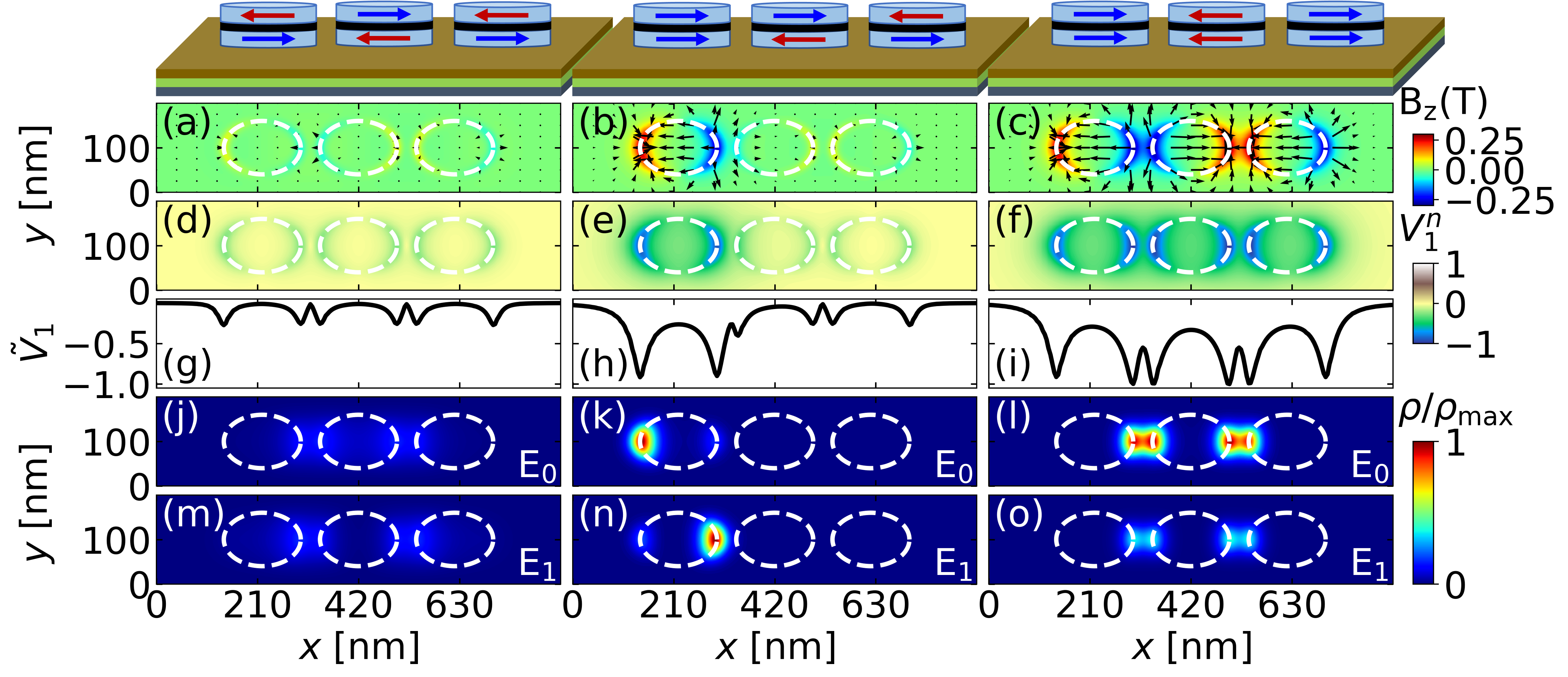}
\caption{Magnetic textures generated by a 3-MJ array and their effects on the quantum states of the 2D system underneath. Each column shows results for the corresponding system configurations sketched at the top of the figure. (a)-(c) Position dependence of the magnetic fringe fields. Colors indicate the strength of the field component along the $z$ direction and the arrows correspond to the in-plane components. (d)-(f) Position dependence of the normalized spin-dependent potential, $V_{1}^n=V_{1}/(\max|V_{1}|)$, where $\max|V_{1}|=2.61$~meV is taken over all positions and configurations. (g)-(i) Variation of the normalized spin-dependent potential along the $x$ direction for $y=0$ [$\tilde{V}_1(x)=V_1^n(x,0)$]. (j)-(l) and (m)-(n) Probability densities (normalized to their maximum value) for the ground and first excited states, respectively.}
\label{fig:rho-3}
\end{figure}

To illustrate how arrays of multiple MJs can generate different sets of quantum states we now consider a chain composed of three MJs. There are 9 different magnetic textures that can be realized from combining the configurations of the MJs. For simplicity we focus on the three possibilities depicted in Fig.~\ref{fig:rho-3} but the analysis can easily be extended to the other 6 combinations. The fringe fields, spin-dependent confining potential, and normalized probability densities for the ground and first excited state are shown in the left, middle, and right columns of Fig.~\ref{fig:rho-3} for the OFF-OFF-OFF, ON-OFF-OFF, and ON-ON-ON configurations, respectively. In the OFF-OFF-OFF configuration the fringe field is weak [see Fig.~\ref{fig:rho-3}(a)] and the spin-dependent potential $V_1$ generates six shallow QDs [see Figs.~\ref{fig:rho-3}(d) and (g)] supporting weakly bounded ground and first excited states with energies $E_0 = -0.02$~meV and $E_1=-0.01$~meV [see Figs.~\ref{fig:rho-3}(j) and (m)], respectively. The ON-OFF-OFF configuration, realized by switching the left MJ, yields an asymmetric potential, where the two leftmost QDs deepen while the other four remain shallow [see Fig.~\ref{fig:rho-3}(e) and (h)]. Due to the asymmetry of the MJ configuration, the strengths of the confining potential of the two leftmost QDs are slightly different. This results in the ground ($E_0 =-1.04$~meV) and first excited ($E_1=-1.02$~meV) states being well localized in the leftmost and second leftmost effective QDs, as shown in Figs.~\ref{fig:rho-3}(k) and (n), respectively.

When the three MJs are in the ON state, the fringe fields become sizable [see Fig.~\ref{fig:rho-3}(c)]. As the effective QDs generated by the fringe field deepen and the overall confinement potential exhibit mirror symmetry with respect to the planes perpendicular to the $x$ and $y$ axes [see Figs.~\ref{fig:rho-3}(f) and (i)], the ground and first excited states become quasi-degenerate, with energies $E_0=-1.426$~meV and $E_1=-1.423$~meV, respectively, and well localized in the regions between the MJ projections on the 2D system [see Figs.~\ref{fig:rho-3}(l) and (o)].

Figure \ref{fig:rho-3} demonstrates that by changing the magnetic configuration of the MJs one can not only tune the energy spectrum in the 2D system but also the localization of the quantum states.

\begin{figure}[t]
\centering
\includegraphics[width=1\linewidth]{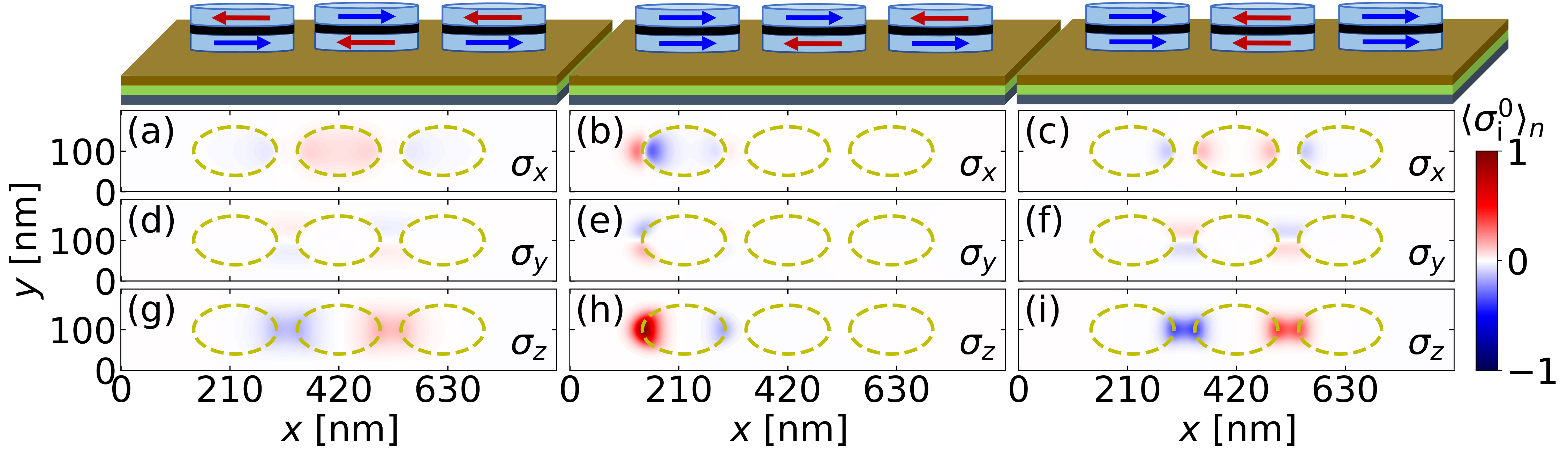}
\caption{Position dependence of the ground-state spin density (normalized to its maximum value) generated by the fringe fields of a 3-MJ array. Each column shows the $x$, $y$, and $z$ components of the spin density for the corresponding system configuration sketched at the top of the figure.}
\label{fig:s-3-0}
\end{figure}

The normalized ground-state spin densities corresponding to the considered magnetic configurations of the 3-MJ array are shown in Fig.~\ref{fig:s-3-0}. Similarly to the single-MJ case, the strength of the spin density is determined by the ground state localization, and its direction follows the local magnetic field orientation. The fringe fields corresponding to the OFF-OFF-OFF and ON-ON-ON configurations obey Eqs.~(\ref{inv-x}) and (\ref{inv-y}) and therefore the symmetry relations in Eqs.~(\ref{h-sim-rx}) -(\ref{s-sim-y}) also hold. However, switching to the ON-OFF-OFF configuration lowers the symmetry and only Eqs.~(\ref{inv-y}), (\ref{h-sim-ry}) and (\ref{s-sim-y}) remain valid.

\begin{figure}[t]
\centering
\includegraphics[width=1\linewidth]{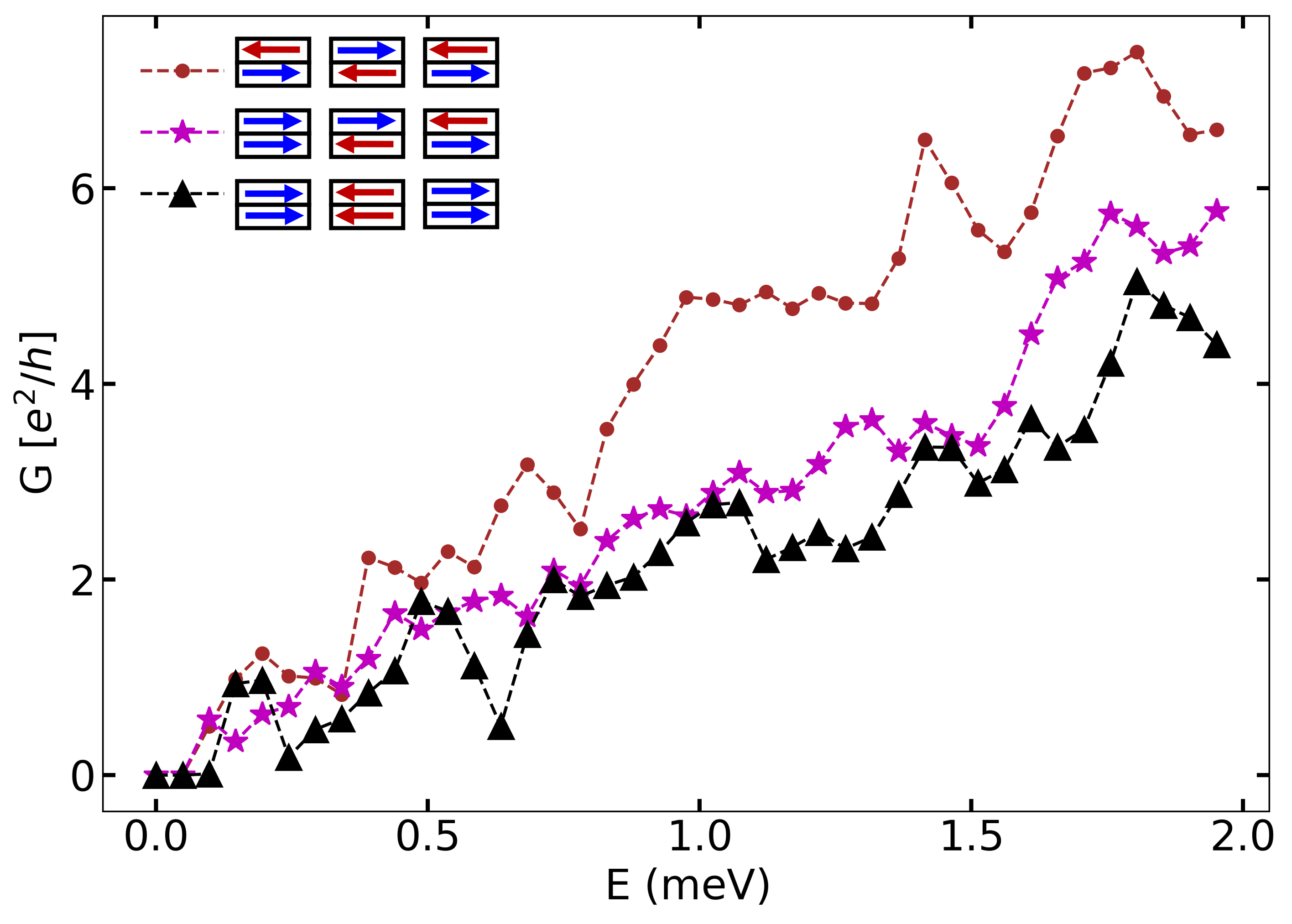}
\caption{Energy dependence of the differential conductance for three different configurations of a 3-MJ array.}
\label{fig:g-3}
\end{figure}

The energy dependence of the differential conductance is shown in Fig.~\ref{fig:g-3} for three configurations of the 3-MJ array. The qualitative behavior is similar to the single-MJ case, with a non-monotonic conductance exhibiting several maxima and minima as the energy varies, and the absence of conductance quantization plateaus. Furthermore, like in the case of a single MJ, in most energy regions, the conductance of the system with the 3-MJ array acquires distinct values for different magnetic configurations of the MJ array. This allows for controlling the transport properties of the 2D system by properly switching individual MJs. 

Information about the MJ array configurations can in principle be extracted by setting an operational energy, at which the conductance differences are sizable, and measuring the conductance through the 2D system. For example, according to Fig.~\ref{fig:g-3}, at $E\approx 1.7$~meV the highest, middle, and lowest values of the measured conductance will indicate that the MJ array is in the OFF-OFF-OFF, ON-OFF-OFF, and ON-ON-ON configurations, respectively. However, the reading can be ambiguous because some configurations (eg., ON-OFF-OFF and OFF-OFF-ON) may lead to the same conductance. This could be alleviated by attaching multiple leads to the 2D system and measuring the conductance between different pairs of leads.

\section{Conclusions}

We have investigated the effects of magnetic textures generated by an array of MJs on the quantum states of a 2D system underneath. The magnetic fringe field created by each MJ is weak when the junction is in the antiparallel (OFF) configuration but becomes sizable when the junction is switched to the parallel (ON) state. This enables local modulations of the collective magnetic texture on the nanometer scale by switching individual MJs. The magnetic texture couples to the motion and spin of the carriers in the 2D system through the kinetic momentum and the Zeeman interaction, respectively. By performing local spin rotations aligning the spin quantization axis to the local magnetic field direction we have shown that the noncollinearity of the magnetic texture generates a non-Abelian field. The non-Abelian field gives rise to a synthetic SOC, whose strength, symmetry, and range can be magnetically tuned. As predicted by our theoretical analysis and demonstrated through numerical simulations, the magnetic textures produced by MJ arrays generate spin-dependent potentials. The spin-dependent potential is attractive for one of the spin species and forms effective QDs that can support bound states in 2D systems with large effective $g$-factor. Since the position and depth of the effective QDs strongly depend on the magnetic configuration of the MJ-array, the energy spectrum and localization of the quantum states in the 2D system can be tuned by changing the state of individual MJs in the array. Therefore, the use of arrays containing multiple MJs could provide a way for designing a large
number of different sets of quantum states, as well as controlling the transitions between them. For the other spin species, the spin-dependent potential acts as potential barriers whose locations and strengths are determined by the MJ-array configuration. Therefore the charge transport through the 2D system can be modulated by reconfiguring the magnetic texture generated by the MJ-array.

\appendix*
\section{Spin-Orbit Coupling Effects}

In the presence of both bulk and structure inversion asymmetries the SOC Hamiltonian can be approximated as,
\begin{equation}\label{Hso}
H_{so}=H_R+H_D,
\end{equation}
where,
\begin{equation}\label{HR}
H_{R}=\frac{\alpha}{\hbar}\left(\pi_y\sigma_x-\pi_x\sigma_y\right),
\end{equation}
is the Rashba SOC and
\begin{equation}\label{HD}
H_{D}=\frac{\beta}{\hbar}\left(\pi_x\sigma_x-\pi_y\sigma_y\right),
\end{equation}
represents the linearized Dresselhaus SOC. The coordinate axes are chosen such that $\mathbf{x}\parallel [100]$ and $\mathbf{y}\parallel [010]$, and $\alpha$ ($\beta$) represents the strength of the Rashba (linearized Dresselhaus) SOC.

The tight-binding representations of the Rashba and Dresselhaus SOCs on the discretized lattice read,
\begin{eqnarray}\label{HR_tb}
H_{R} &=& \frac{\alpha}{\hbar\; a\;t}\sum_{\mathbf{n},\sigma,\sigma^\prime}\{ i\; t_{\mathbf{n},\mathbf{n}+\mathbf{e}_y}S^x_{\sigma\sigma^\prime}|\mathbf{n},\sigma\rangle\langle \mathbf{n}+\textbf{e}_y,\sigma^\prime|\nonumber\\ 
&-& i\; t_{\mathbf{n},\mathbf{n}+\mathbf{e}_x}S^y_{\sigma\sigma^\prime}|\textbf{n},\sigma\rangle\langle \textbf{n}+\textbf{e}_x,\sigma^\prime| + \mathrm{H.c.} \}
\end{eqnarray}
and
\begin{eqnarray}\label{HD_tb}
H_{D} &=& \frac{\beta}{\hbar\; a\;t}\sum_{\mathbf{n},\sigma,\sigma^\prime}\{ i\; t_{\mathbf{n},\mathbf{n}+\mathbf{e}_x}S^x_{\sigma\sigma^\prime}|\mathbf{n},\sigma\rangle\langle \mathbf{n}+\textbf{e}_x,\sigma^\prime|\nonumber\\ 
&-& i\; t_{\mathbf{n},\mathbf{n}+\mathbf{e}_y}S^y_{\sigma\sigma^\prime}|\textbf{n},\sigma\rangle\langle \textbf{n}+\textbf{e}_y,\sigma^\prime| + \mathrm{H.c.} \}
\end{eqnarray}

Numerical calculations including the SOC effects were performed for CdMnTe with $\alpha=0.33$~meV~nm and $\beta=0.46$~meV~nm \cite{baboux2013:PRB}. The results did not show major changes in the energy spectrum and localization of the quantum states. This is not surprising because the large effective $g$-factor  in CdMnTe leads to field-induced effective QDs with confining potentials considerably stronger than the SOC strength. However, this may not be the case in materials possessing stronger SOC and smaller $g$-factors (e.g., InGaAs/InAs quantum wells).

The SOC Hamiltonians satisfy the following commutation,
\begin{equation}\label{hso-comm}
    [H_R,\mathcal{R}_x\mathcal{R}_{sx}]=[H_R,\mathcal{R}_y\mathcal{R}_{sy}]=0,
\end{equation}
and anticommutation,
\begin{equation}\label{hso-anticomm}
    \{H_D,\mathcal{R}_x\mathcal{R}_{sx}\}=\{H_D,\mathcal{R}_y\mathcal{R}_{sy}\}=0,
\end{equation}
relations. Therefore, SOC can lead to changes in the overall symmetry of the system. For example, for the case of a single MJ, the symmetry relations in Eqs.~(\ref{h-sim-rx}) and (\ref{s-sim-x}) remain valid in the presence of Rashba SOC but no longer hold when Dresselhaus SOC is present. Conversely, Dresselhaus SOC preserves Eqs.(\ref{h-sim-ry}) and (\ref{s-sim-y}) but Rashba SOC breaks them.

Although in the example discussed above the combined space and spin mirror symmetries are broken when Rashba and Dresselhaus SOCs coexist, there are other MJ arrays with magnetic configurations in which certain symmetries combining space and spin mirror operations are preserved even in the presence of both Rashba and Dresselhaus SOCs. As an example, consider a 2-MJ array with the two MJs in the ON state and opposite magnetization directions (e.g., like the two leftmost MJs in the array sketched in the right column of Fig.~\ref{fig:rho-3}). If the MJs are located symmetrically with respect to the center of the 2D system, the generated fringe field obeys the relations,
\begin{equation}\label{rx-sim-2mj}
    \{\mathcal{R}_x,B_x\}=0\;,\;[\mathcal{R}_x,B_y]=[\mathcal{R}_x,B_y]=0,
\end{equation}
and,
\begin{equation}\label{ry-sim-2mj}
[\mathcal{R}_y,B_x]=[\mathcal{R}_y,B_z]=0\;,\;\{\mathcal{R}_y,B_y\}=0.
\end{equation}
It then follows that the total Hamiltonian, including the kinetic energy ($H_0$), the Zeeman interaction with the fringe fields ($H_Z$) and both Rashba and Dresselhaus SOCs obeys the commutation relation,
\begin{equation}\label{rx-ry-rsz}
    [\mathcal{R}_x \mathcal{R}_y \mathcal{R}_{sz},H_0+H_Z+H_R+H_D]=0.
\end{equation}
As a direct consequence of Eq.~(\ref{rx-ry-rsz}), the spin-density components exhibit the following symmetry properties,
\begin{eqnarray}\label{s-sim-x-y-sz}
    \langle\sigma_x\rangle(x,y)&=&-\langle\sigma_x\rangle(-x,-y)\;,\nonumber\\
    \langle\sigma_y\rangle(x,y)&=&-\langle\sigma_y\rangle(-x,-y)\;,\nonumber\\
    \langle\sigma_z\rangle(x,y)&=&\langle\sigma_z\rangle(-x,-y).
\end{eqnarray}

\bibliographystyle{apsrev4-2}

\begin{thebibliography}{34}
\bibitem [{Not()}]{Note:1}%
	\BibitemOpen
	\href@noop {} {} \ \bibinfo {note} {The second term in
		Eq.~(\ref{v-sigma-1}) dominates in regions where the fringe field is sizable.
		In regions where the fringe field is negligibly small, the first contribution
		can be gauged out.}\BibitemShut {Stop}%
\bibitem [{Not()}]{Note:2}%
	\BibitemOpen
	\href@noop {} {} \ \bibinfo {note} {Note that
		$V_\sigma(\mathbf{r})<0$ over a finite region of the 2D system is a necessary
		but not sufficient condition for the emergence of bound states.}\BibitemShut
	{Stop}%
\end{thebibliography}
%

\end{document}